\documentclass[aps, prb, reprint, showpacs, superscriptaddress]{revtex4-1}
\pdfoutput=1
\usepackage{graphicx}% Include figure files
\usepackage{dcolumn}% Align table columns on decimal point
\usepackage{bm}% bold math
\usepackage[colorlinks=true, linkcolor=red, citecolor=blue, urlcolor=blue, linktoc=page, bookmarks=false, pdfstartview={FitH}, pdfborder={0 0 0.0 [3 3]}]{hyperref}
\usepackage{url}
\usepackage{color}
\usepackage{amsmath}
\usepackage{natbib}
\usepackage{dcolumn}% Align table columns on decimal point
\newcolumntype{.}{D{.}{.}{-1}}

\begin{document}
\title{Role of Te in the low dimensional multiferroic material FeTe$_\text{2}$O$_\text{5}$Br}
%_________AUTHOR_________ 
\author{Jayita Chakraborty}
\affiliation{Department of Solid State Physics, Indian Association for the Cultivation of Science, Jadavpur, Kolkata 700032, India}

\author{Nirmal Ganguli}
\affiliation{Faculty of Science and Technology and MESA$^+$ Institute for Nanotechnology, University of Twente, P.O. Box 217, 7500 AE Enschede, The Netherlands}

\author{Tanusri Saha-Dasgupta}
\affiliation{S. N. Bose National Center for Basic Sciences, JD-III, Salt Lake City, Kolkata 700098, India}

\author{Indra Dasgupta}
\email{sspid@iacs.res.in}
\affiliation{Department of Solid State Physics, Indian Association for the Cultivation of Science, Jadavpur, Kolkata 700032, India}

\date{\today}% It is always \today, today,

\begin{abstract}
%We have employed
Using  first principles density functional calculations,  we study the electronic structure of the low-dimensional multiferroic compound  FeTe$_2$O$_5$Br to investigate the origin of the magnetoelectric (ME) effect  and the role of Te ions  in this system. We find that without magnetism  even in the presence of   Te-5$s$ lone pairs, the system remains centrosymmetric  due to the antipolar orientation of the lone pairs. Our study shows that the exchange striction within the Fe tetramers as well as between them is  responsible for the ME effect in FeTe$_2$O$_5$Br. We also find that the  Te$^{4+}$ ions play an important role in the inter-tetramer exchange striction as well as contribute to  the electric polarization in  FeTe$_2$O$_5$Br, once the polarization is triggered by the magnetic ordering.
\end{abstract}
\pacs{71.20.-b, 75.30.Et, 75.80.+q, 77.80.-e.}
%\pacs{75.10.Jm, 75.30.Et, 71.20.-b}
%\pacs{75.10.Jm., 71.20.−b, 75.30.Et.}% PACS, the Physics and Astronomy
                              %display desird
\maketitle
%===================================================
\section{Introduction}
Multiferroic  materials with the simultaneous presence of  ferroelectricity and magnetism, have been the focus of  attention in recent times.\cite{kimuraNature, nicolaScience} Based on the microscopic origin of ferroelectricity (FE) multiferroic materials can be classified into two different classes namely type-I(proper)  and type-II(improper)  multiferroic materials. In type-I multiferroics, ferroelectricity and magnetism stem from an independent origin and  the coupling between magnetism and ferroelectricity is usually weak.  In these materials, ferroelectricity typically appear at higher temperatures than magnetism, and the magnitude of spontaneous electric polarization ($P$) is often  large ($\sim$ 10-100 $\mu$C/cm$^2$). One possible mechanism  for ferroelectricity in type-I multiferroic material is lone-pair driven. It is well known that cations containing  highly polarizable 5$s$ or 6$s$ lone pairs of valence electrons have a strong tendency to break the local inversion symmetry of the crystal. This lone-pair driven mechanism was identified as the source of ferroelectric instability in ${\rm BiFeO_3}$.\cite{RavindranPRB2006} In contrast, type-II multiferroics where ferroelectricity may arise due to a particular  kind of magnetic ordering that breaks the inversion symmetry, are more interesting from an  application point of view due to the strong coupling between  magnetism and FE.\cite{khomskii2006, cheong2007} However, the magnitude of electric polarization in these materials is  usually very small ($\sim$ 10$^{-2}$ $\mu$C/cm$^2$).  For type-II multiferroics, non symmetric lattice distortion and ferroelectric order may be induced through exchange striction,\cite{LeePRB2011, cheong} spin current mechanism \cite{knb} or inverse Dzyaloshinskii-Moriya  (DM) interactions. \cite{SergienkoPRB} In particular, the exchange striction is considered to induce  ferroelectricity in some  collinear antiferromagnets such as HoMnO$_3$,\cite{LeePRB2011} $\rm{ Ca_3CoMnO_6}$.\cite{cheong, Wu2009, whangboccmo} While strong coupling between the magnetic and ferroelectric order parameters  makes them attractive but their real applications have been restricted by the small magnitude of the  polarization values. A possible way to overcome this difficulty could be to combine the best features of type-I and type-II multiferroics. In this context, the transition metal (TM) selenium (Se) and tellurium (Te)  oxihalides may offer an attractive possibility as they exhibit exotic magnetic properties driven by the geometric frustration in low dimensions and also contain stereochemically active lone pair in p element cations Te$^{4+}$, and  Se$^{4+}$ that may result in lone pair driven ferroelectricity as in the case of type-I mutiferroics. 
% Therefore such systems offer an interesting possibility of combining the features of type-I (presence of lone pair) and type II ferroelectricity (novel magnetic order in low dimensions). 
Interestingly some of  these systems exhibit multiferroic behavior. An example of such a system is FeTe$_2$O$_5$Br. 
It adopts a layered structure, where individual layers consist of geometrically frustrated iron tetramer units [Fe$_4$O$_{16}$]  linked by the [Te$_4$O$_{10}$Br$_2$]$^{6-}$ groups.\cite{beckerjacs2006} However the structure remains centrosymmetric even in the presence of Te-$5s^2$ lone pairs.  The high temperature fit to the susceptibility data shows  negative Curie-Weiss temperature ($\theta_{\text{CW}} = -98$ K), indicating  strong antiferromagnetic interactions between the Fe$^{3+}$$(d^5)$ ions.\cite{beckerjacs2006} The system develops long range magnetic order at a considerable low temperature $T_{N1} = 11$ K, followed by a second magnetic transition at $T_{N2} = 10.5$ K.\cite{pregeljprb2010} The first transition at  $T_{N1}$ is a paramagnetic to a high-temperature incommensurate  magnetic state (HT-IC) with a constant wave vector $q_{IC1} = ( 0.5, 0.466, 0.0)$ and is immediately followed by another transition at $T_{N2} = 10.5$ K  into the low-temperature incommensurate (LT-IC) multiferroic state. The amplitude modulated magnetic order in the LT-IC phase is described with the wave vector $q = (0.5, 0.463, 0)$  and  concomitantly with the magnetic order a ferroelectric polarization ($P=8$ $\mu$C/m$^2$) is induced perpendicular to $q$ and the direction of the Fe$^{3+}$ moments.\cite{pregeljprl2009} As the polarization is found to be triggered by magnetic ordering, the small value of the polarization provide direct evidence that FeTe$_2$O$_5$Br is an example of type~II (improper) multiferroic contrary to the original expectation.  
A recent study\cite{Pregelj2013} on the magnetic ordering in the HT-IC phase of FeTe$_2$O$_5$Br showed that while the inversion symmetry is already broken in HT-IC phase, the ferroelectricity is only realized in the LT-IC phase. The difference in the orientation of the magnetic moments and phase shift of the amplitude modulated waves between the two magnetic structures is suggested to be responsible for the realization of ferroelectricity in the LT-IC phase.
%The origin of the electric polarization is ascribed to the exchange striction of the interchain interactions. A recent study on the magnetic ordering in the HT-IC phase of FeTe$_2$O$_5$Br,  showed that  although all crystal symmetries are broken already in the HT-IC phase but the two magnetic phases differ in the orientation of the magnetic moments and phase shifts between the amplitude modulated  waves. 
In addition, there is evidence of minute displacements of the Te$^{4+}$ ions  in the  LT-IC phase and these subtle displacements  may  be important for the electric polarization in this phase.\cite{Pregelj2013} In view of the above, it is suggested that polarization is possibly driven by exchange striction on the inter-chain bond containing  highly polarizable Te lone pair electrons. In the search for a suitable spin Hamiltonian, magnetic susceptibility was analyzed by various groups. An early report suggested that magnetic susceptibility can be explained by considering the dominant interactions within the Fe tetramers.\cite{beckerjacs2006}  A recent study, however, shows that  the system should be described as a system of alternating antiferromagnetic $S = 5/2$ chains with strong Fe-O-Te-O-Fe bridges weakly coupled by two-dimensional frustrated interactions.\cite{henaarxiv}

The preceding discussion suggests that it  will be important to clarify the role of Te ions in the  multiferroic property of  FeTe$_2$O$_5$Br. In particular, it will be interesting to understand the interplay of  magnetic interaction and the activity of the Te$^{4+}$ lone pairs and eventually their combined role in the ferroelectric polarization.  In the present paper we have examined this issue in details using {\em ab initio} electronic structure calculations. The remainder of the paper is organized as follows. In Sec II we  describe the crystal structure along with  the computational details. Section III is devoted to a detailed discussion of our results on the electronic structure calculations. Finally, a summary and conclusions are given in Sec IV. 

\section{Crystal Structure and Computational Details}
\begin{figure}[h]
\centering
\includegraphics[scale=.26]{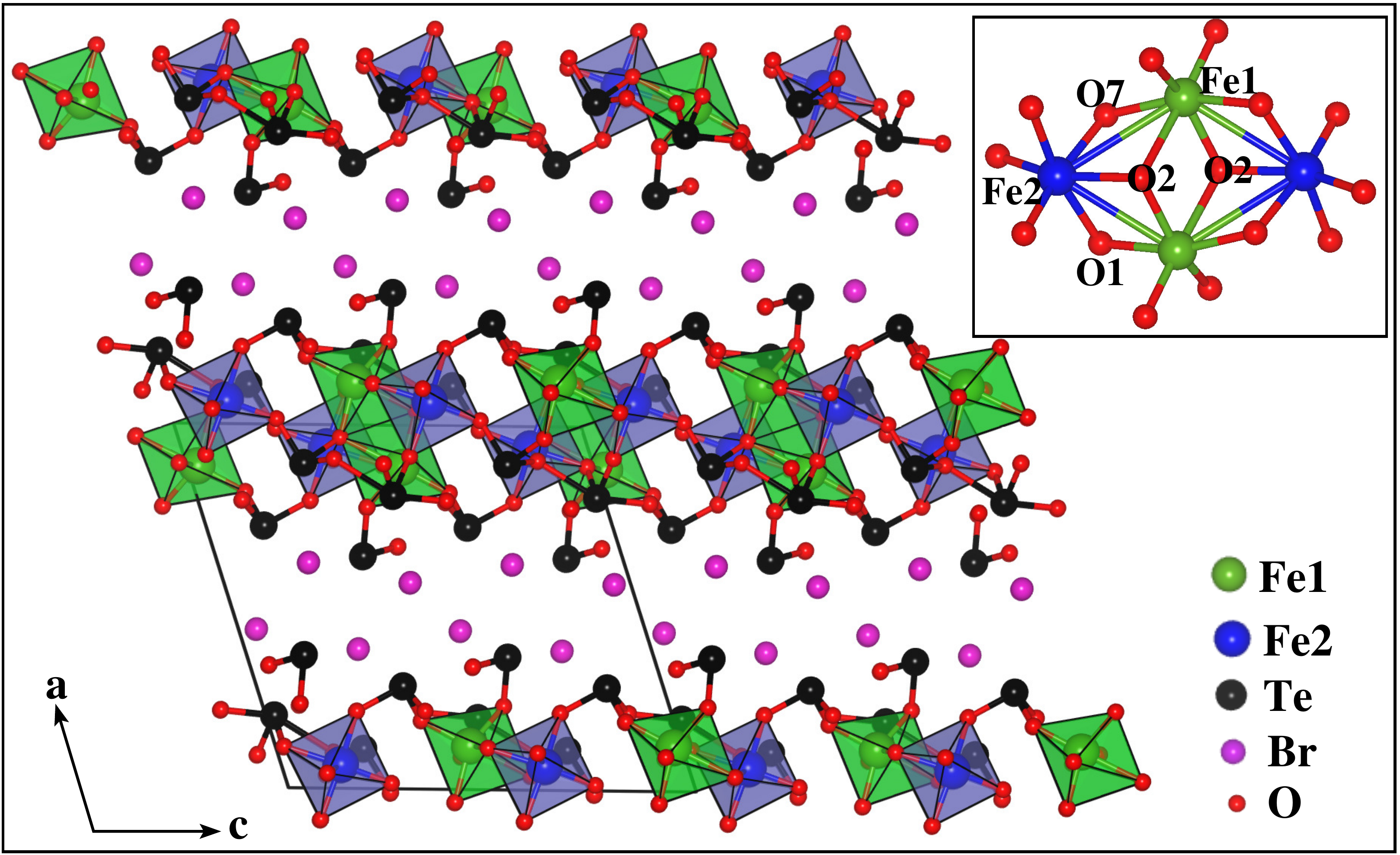}
\caption{\label{fig:structure}Layered structure of FeTe$_2$O$_5$Br. The inset shows one tetramer unit.}
\end{figure}

FeTe$_2$O$_5$Br crystallizes in the monoclinic space group $P21/c$. The crystallographic unit cell has an inversion center. The lattice parameters for FeTe$_2$O$_5$Br are $a=13.396$ \AA, $b=6.597$ ~\AA, $c=14.289$ \AA~and $\beta=108.12^{\circ}$.\cite{beckerjacs2006} The unit cell  (depicted in Fig.~\ref{fig:structure}) contains 72 atoms.
%.............Fig.1 ............
There are two crystallographycally inequivatent  Fe$^{3+}$ ions  in the structure which are in distorted [${\rm FeO_6}$] octahedral environment. Four such octahedra share their edges with each other and form a [Fe$_4$O$_{16}$] iron tetramer cluster (see the inset of Fig.~\ref{fig:structure}). These iron tetramers are  linked by [Te$_4$O$_{10}$Br$_2$]$^{6-}$ units forming a layered structure in the $bc$ plane. The layers are weakly connected via van der Waals forces as they stack along the monoclinic $a$ axis.

The first principles density functional  theory (DFT) calculations have been performed using the plane wave based projector augmented wave (PAW) \cite{blochl} method  as implemented in the Vienna {\em ab initio} simulation package (VASP).\cite{vasp} We have used local density approximation (LDA) to the exchange correlation functional. The localized Fe-$d$ states were treated in the framework of local spin-density approximation (LSDA)+$U$ method,\cite{dudarev1998} where calculations were done for several values of $U_{eff}$= $U$-$J=$ in the range $0$(LDA) $-5$ eV.   The calculations for the unit cell were performed with a ($4\times8\times4$)  $\Gamma$ centered $k$ point mesh and 550 eV as plane wave cut off energy. In order to simulate the magnetic structure we have neglected the amplitude modulation and have approximated the incommensurate wave vector q$\sim$ (0.5, 0.463, 0) by a commensurate one (0.5, 0.5, 0) and have generated a supercell ($2\times2\times1$) of the original unit cell containing 288 atoms. For the calculations with the supercell, a plane-wave cutoff energy of 500 eV was used along with a   ($1\times2\times2$)  $\Gamma$ centered $k$ point mesh. All structural relaxations were carried out till Hellman-Feynman forces became less than 0.01  eV/\AA.

 For the derivation of the low energy model Hamiltonian and identification of various exchange paths we have employed N-th order muffin-tin orbital (NMTO) downfolding method.\cite{Saha-Dasgupta,Tank} The NMTO downfolding method is an efficient {\em ab-initio} scheme to construct low energy, few band tight-binding model Hamiltonian. The low energy model Hamiltonian is constructed by energy selective downfolding method, where high energy degrees of freedom are integrated out from the all orbital LDA calculations. The fourier transform of the resulting low energy Hamiltonian yields the effective hopping parameters which can be utilized to identify the dominant exchange paths. %(Jayita put the NMTO references here)}

%.............Fig2............
\begin{figure}
\centering
\includegraphics[scale=.45]{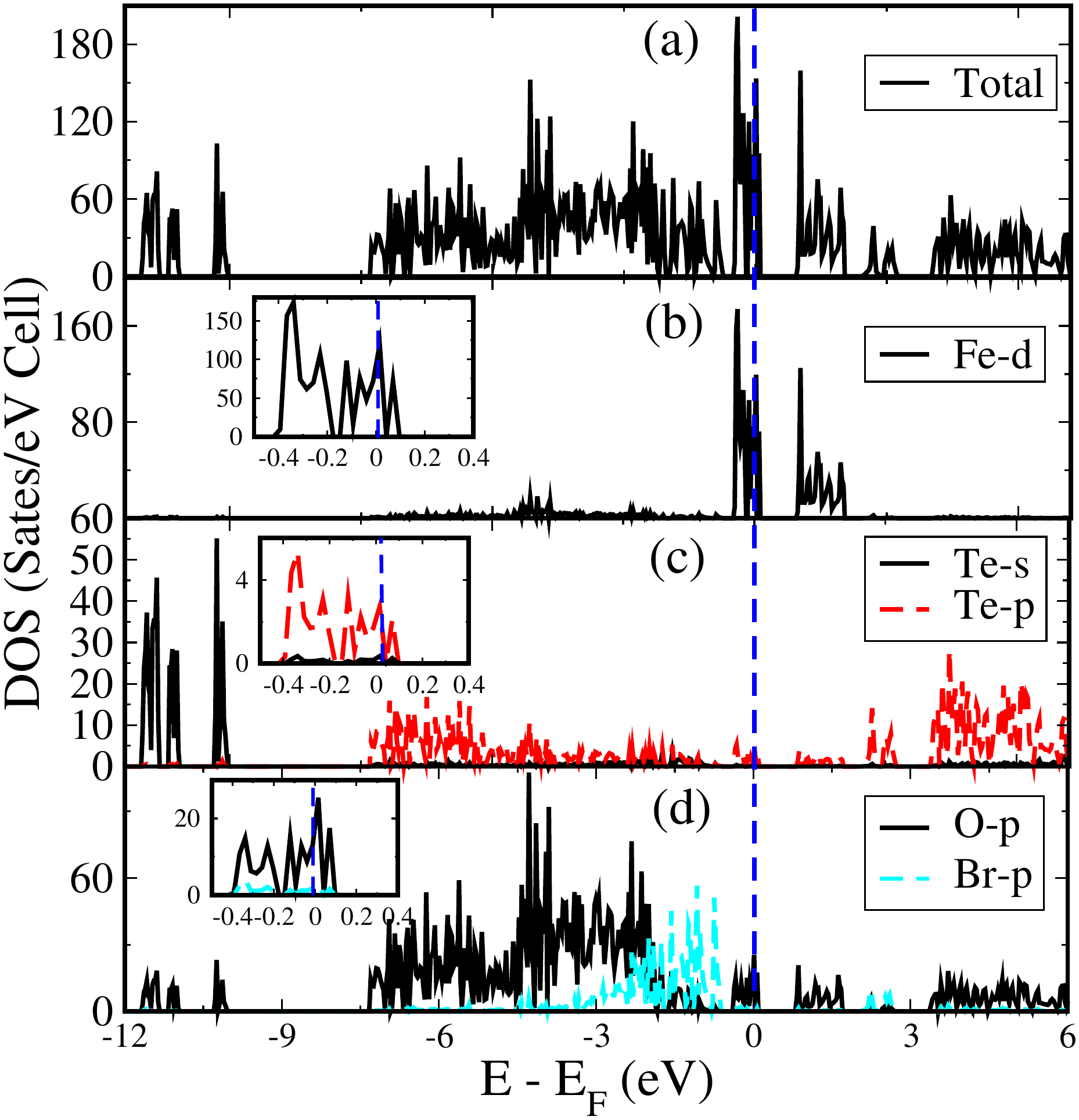}
\caption{\label{fig:parados}The non-spin-polarized density of states for  ${\rm FeTe_2 O_5Br}$. (a) the total DOS, orbital-projected density of states for (b) Fe-$d$ , (c) Te-$s$ and Te-$p$, and (d) O-$p$ and Br$p$.  Insets show the orbital characters near the Fermi level.}
\end{figure} 
%..................................................
\section {Results and Discussions}
\subsection{Non-spin polarized calculation}
To begin with we have investigated the electronic structure of FeTe$_2$O$_5$Br without magnetic order.  The non-spin polarized total and partial density of states are shown in Fig.~\ref{fig:parados}. The density of states (DOS) is consistent with the Fe$^{3+}$Te$^{4+}_2$O$^{2-}_5$Br$^{1-}$  nominal ionic formula for the system. Fig.~\ref{fig:parados}, reveals that O-$p$ and Br-$p$ states are completely occupied while the Fermi level ($E_{F}$) is dominated by the Fe-$d$ states.  The occupied Te-5$s$ states lie far below the $E_F$. The empty Te-5$p$ states lie above the Fermi level spreading on an energy range  2 - 6 eV with respect to the Fermi level. There is a significant admixture of Te-5$s$ and Te-5$p$ states with the O-$p$ states, which suggests the hybridization between Te and O, which in turn hybridizes with Fe-$d$ states crossing the Fermi level (see insets in Fig.~\ref{fig:parados} ).

\begin{figure}
\centering
\includegraphics[scale=.35]{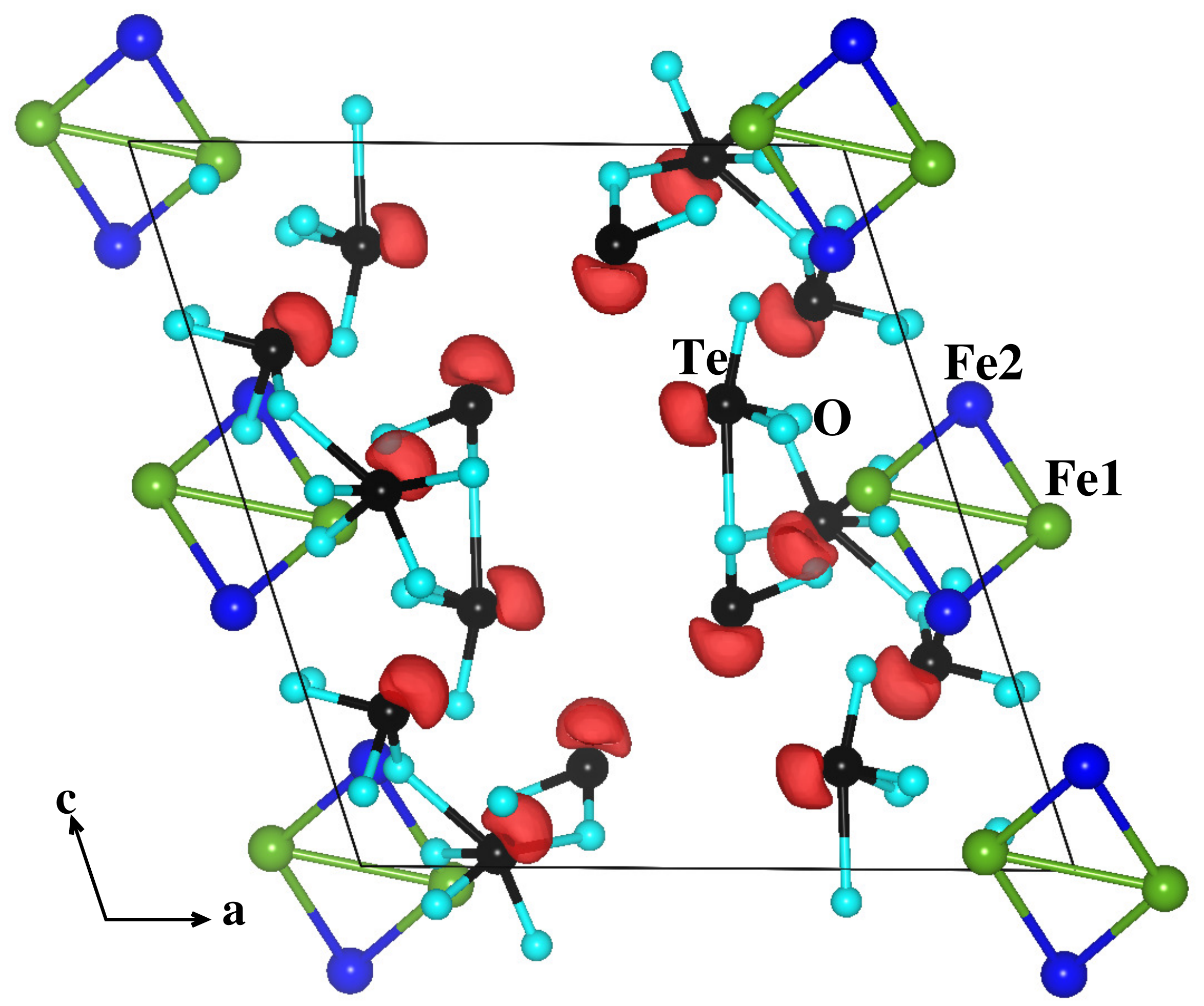}
\caption{\label{fig:elfplot}Electron localization function within a unit cell. The isosurfaces are visualized for a value of 0.9.}
\end{figure}

The presence of Te in 4+ oxidation state suggests the possibility of the stereochemical activity of Te lone pair formed from $5s^2$ electrons.  In order to visualize the lone pairs arising from 5$s^2$ electrons of Te$^{4+}$ ions, we have calculated the electron localization function (ELF).\cite{elf-nature,savinAng1997} The ELF is defined as follows:
\begin{align}
{\text{ELF}} &= \left[1+ \left(\frac{D}{D_h}\right)^2\right]^{-1} \\
{\text{where }} 
D &= \frac{1}{2}\sum_i|\nabla \phi_i|^2- \frac{1}{8}\frac{|\nabla \rho|^2}{\rho} {\text{ and}} \nonumber \\
D_h  &=  \frac{3}{10}(3\pi^2)^{5/3}\rho^{5/3}.  \nonumber 
 \end{align}
$\rho$ is the electron density and $\phi_i$ are the Kohn Sham wave functions. The ELF is defined in such a way that its value lies between 0 and 1. The values are close to 1 when in the vicinity of one electron, no other electron  with the same spin may be found. For instance this would occur in bonding pairs or {\em lone pairs}.\cite{Peters1992}  From the plot of the electron localization function, in the experimental high temperature centrosymmetric structure\cite{beckerjacs2006} displayed in Fig.~\ref{fig:elfplot}, we find that the electron density  around Te is asymmetric and form a usual lobe shape arising from the $5s$ lone pair of Te.  It has been pointed out by Watson and Parker that the hybridization with anion $p$ orbitals (oxygen 2$p$) plays an  important role in the formation of asymmetric lobe shaped isosurface of electron localization function for sterically active lone pairs.\cite{Watson1999} We gather from the DOS shown in Fig.~\ref{fig:parados} that the occupied Te $s$ and O $p$ orbitals hybridize to form a pair of occupied bonding and antibonding states. This Te-5$s$ -O-2$p$ mixed state further hybridizes with empty Te-5$p$ states.  As a consequence both the Te-5$s$ and Te-5$p$ states are involved in the formation of the asymmetric electron distribution where empty Te 5$p$ orbitals are able to interact due to  the presence of Te-$s$ - O-$p$ occupied antibonding states. This  emphasizes the importance of the O-$p$ states in the formation of lone pairs.

%%%%%---------Fig3...........

 In order to quantify the hybridization, we have calculated the hybridization index defined as follows:
%=====================
\begin{eqnarray}
	H_{I-l,J-l^{\prime}} &=& \sum_k \left( \sum_i h_{I-l,J-l^{\prime}} ^{i,k} \right) \times {\rm weight}(k), \nonumber \\
	{\rm where~} h_{I-l,J-l^{\prime}} ^{i,k} &=& \sum_{I,J,m,m^{\prime}} w_{lm,i,k}^{(I)} w_{l^{\prime}m^{\prime},i,k}^{(J)}; \nonumber
\end{eqnarray}
%======================
$ w_{lm,i,k}^{(I)}$ are the coefficients in the spherical harmonic decomposition of the local (partial) charge, associated with $i^{th}$ Kohn-Sham orbital,\cite{HakkinenPRL02} around $I^{th}$ atom. $l,m$ indicates the orbital and the magnetic quantum numbers respectively, $I$ and $J$ are atom indices: $I \in$ \{Te atoms\} and $J \in$ \{O atoms\}, $i$ and $k$ stand for band index and $k$-points respectively.  Weight($k$) is  the weight on each $k$-point in the irreducible Brillouin zone necessary for the integration. Our calculations find that the hybridization index between Te-$p$ and O-$p$ is 6.13 and that between Te-$s$ and O-$p$ is 3.80 for the experimental structure without any magnetism indicating sizable hybridization between Te and O. It is interesting to note that these lone pairs however do not promote structural distortion and the structure remains centrosymmetric,  as the pair of lobes are arranged in an opposite manner, resulting canceling polarization of the structure, as is evident from Fig.~\ref{fig:elfplot}.

\subsection{Spin polarized calculation}
\begin{figure}
\centering
\includegraphics[scale=.42]{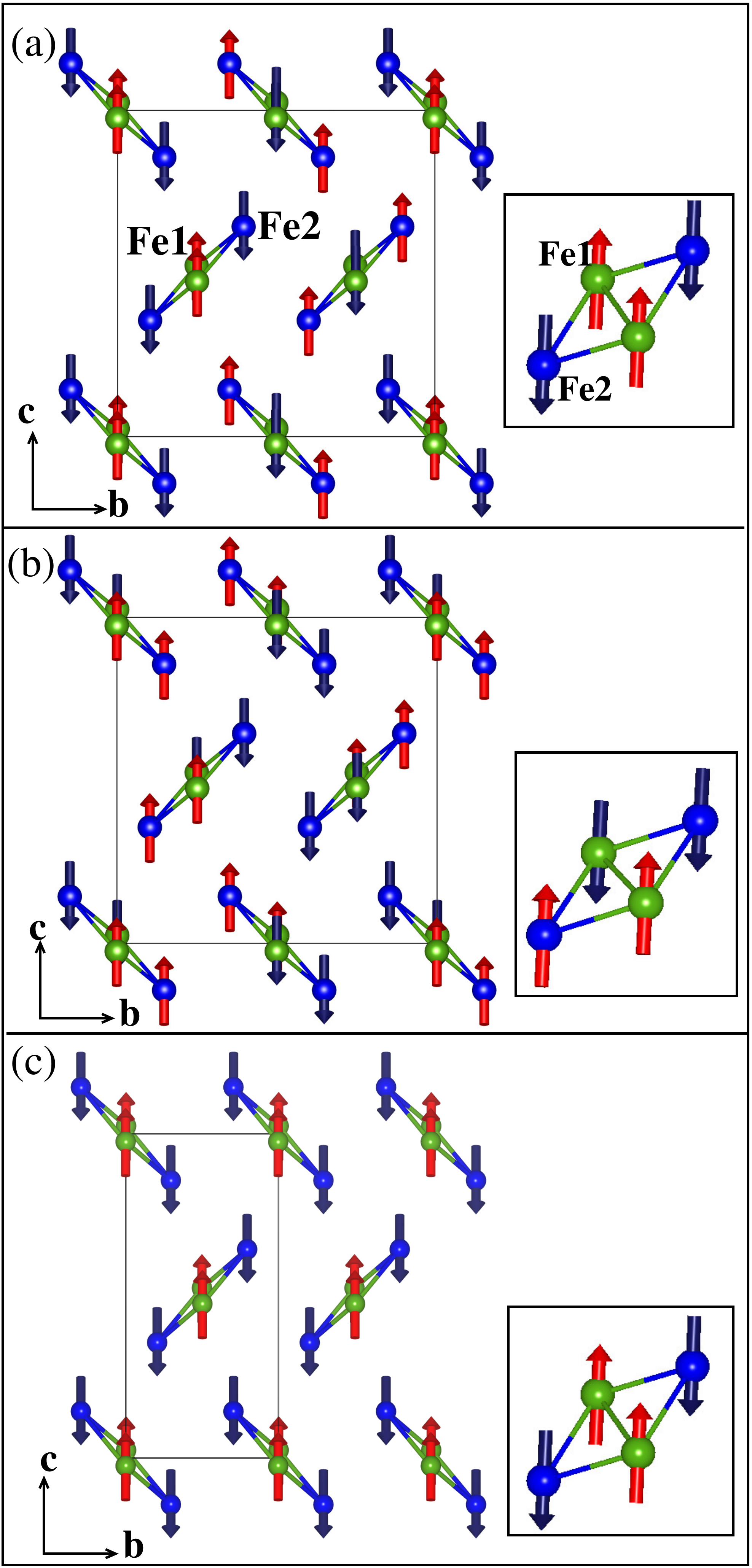}
\caption{\label{fig:afmconfig}Various antiferromagnetic configurations (a) AFM1, (b) AFM2, and (c) AFM3.}
\end{figure}

We  next consider magnetism and its impact on the crystal structure and ferroelectric polarization. In order to simulate the low temperature  magnetic order found in the LT-IC phase we have made a (2$\times$2$\times$1) supercell  which contains 288 atoms. As mentioned before, in our calculation we have neglected the amplitude modulation. We consider  various antiferromagnetic (AFM) configurations (see Fig.~\ref{fig:afmconfig}), depending on the arrangement of Fe spins within each tetramer as well as between the neighboring tetramers. In AFM1  configuration, not only  Fe1 spins aligned antiparallel to Fe2 within each tetramer (see inset of Fig.~\ref{fig:afmconfig}(a)) but also each tetramer is  antiferromagnetically coupled along the $a$ and $b$ directions leading to $q = (0.5, 0.5, 0)$.  The AFM2 configuration differs from the AFM1 configuration only in the arrangement of spins within each tetramer (see inset of Fig.~\ref{fig:afmconfig}(b)) where a pair of Fe1 spins in a tetramer are antiparallel and the same is true for a pair of Fe2 spins. Finally in  AFM3 configuration, the arrangement of Fe1 and Fe2 spins in each tetramer is identical to AFM1 but the tetramers are coupled ferromagnetically along $a$, $b$ and $c$ directions leading to $q = (0, 0, 0)$. The results of our calculations are displayed in Table~\ref{tab:magnetic}.
%In AFM3 configuration, the arrangement of Fe1 and Fe2 spins in each tetramer is identical to AFM1 but $q = (0, 0, 0)$. Finally in  AFM2 configuration in each tetramer a pair of Fe1 spins and Fe2 spins are antiparallel and tetramers are arranged such that $q = (0.5, 0.5, 0)$. The results of our calculations are displayed in Table~\ref{tab:magnetic}.
The results reveal that among the magnetic configurations considered here, AFM1 has the lowest energy. All magnetic states are found to be insulating and the magnetic moment at the Fe site  is $m_{\text{Fe}} \sim 4.2 \mu_B$. The rest of the moments are at oxygen  ($m_{\text{O}} \sim .13 \mu_B$) and bromine ($m_{\text{Br}} \sim .09 \mu_B$) sites, arising due to Fe-O and Fe-Br hybridization effect.

\begin{figure}
\centering
\includegraphics[scale=.35]{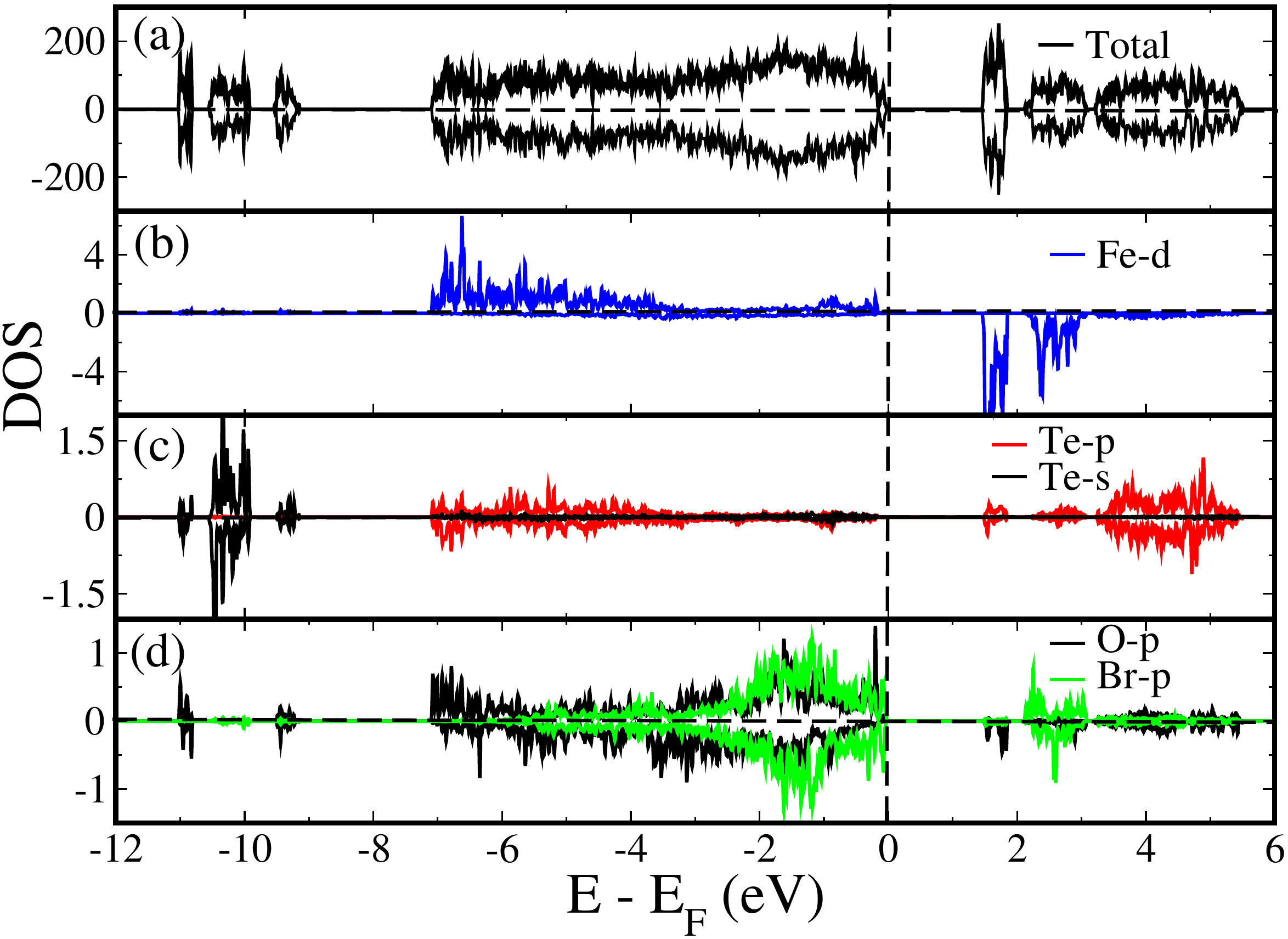}
\caption{\label{fig:pdosafm}The density of states for ${\rm FeTe_2O_5Br}$ in AFM1 configuration with experimental structure. (a) Total DOS (states$/$eV cell). Orbital projected DOS (States$/$eV atom) for (b) Fe-$d$, (c) Te-$s$ and Te-$p$, and (d) O-$p$ and Br-$p$ states.}
\end{figure}

The total density of states as well as its projection onto different atomic orbitals  for AFM1 phase are shown Fig.~\ref{fig:pdosafm}(a-d).  Focusing on  Fig.~\ref{fig:pdosafm}(b), we find that  Fe-$d$ states in the majority spin chanel are completely occupied while the minority states are completely empty, which is consistent with the ${\rm Fe^{3+}}$ valence state of Fe with a $3d^5$ configuration. Such a half filled configuration promotes the AFM order. 

%%--------Fig4...........
%%---------Fig5................

\begin{table*}
\begin{center}
\caption{\label{tab:magnetic} The relative energies, magnetic moments, band gap for different magnetic configurations are listed here.}
\begin{ruledtabular}
\begin{tabular}{c........}
Magnetic & \multicolumn{4}{c}{$U_{eff}=3$~eV} & \multicolumn{4}{c}{ $U_{eff}=5$~eV} \\
\cline{2-5} \cline{6-9}
Config. & \multicolumn{1}{c}{$\Delta E$} & \multicolumn{1}{c}{Band gap} &  \multicolumn{1}{c}{$m_{Fe1}$} &  \multicolumn{1}{c}{$m_{Fe2}$} & \multicolumn{1}{c}{$\Delta E$} & \multicolumn{1}{c}{Band gap} &  \multicolumn{1}{c}{$m_{Fe1}$} &  \multicolumn{1}{c}{$m_{Fe2}$} \\
& \multicolumn{1}{c}{(meV)} & \multicolumn{1}{c}{(eV)} & \multicolumn{1}{c}{($\mu_B$)} & \multicolumn{1}{c}{($\mu_B$)} & \multicolumn{1}{c}{(meV)} & \multicolumn{1}{c}{(eV)} & \multicolumn{1}{c}{($\mu_B$)} & \multicolumn{1}{c}{($\mu_B$)} \\
\hline
FM &   49.7  & 1.3 & 4.1 & 4.1 & 34.1 & 1.5 & 4.2 & 4.2\\
AFM1 &  0.0 & 1.2  & 4.1 & 4.1 &  0.0  & 1.6  & 4.2  & 4.2\\
AFM2 &  15.7 & 1.4 & 4.1 & 4.1 &  10.1 & 1.6 & 4.2 & 4.2\\
AFM3 & 9.0 & 1.5  & 4.1 & 4.1 & 5.8 & 1.7 & 4.2  & 4.2\\
\end{tabular}
\end{ruledtabular}
\end{center}
\end{table*}
%=============================================

\begin{figure}
\centering
\includegraphics[scale=1.15]{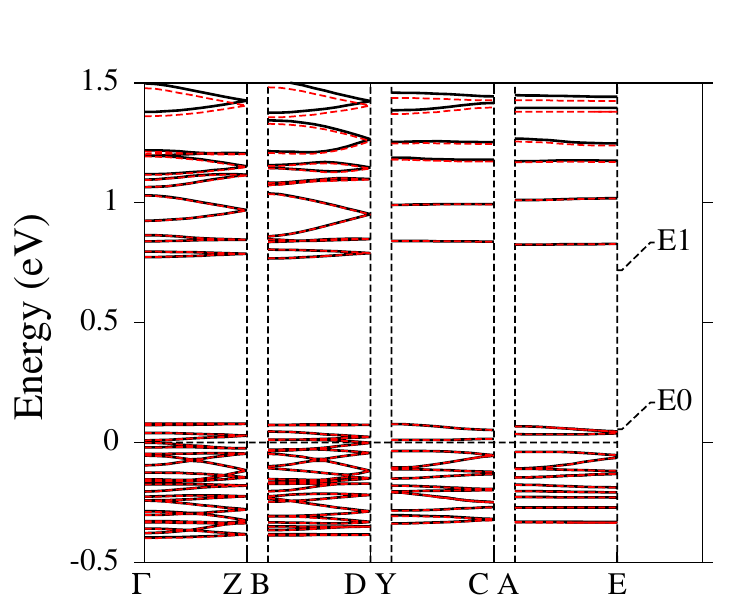}
\caption{\label{fig:nmtoplot}Downfolded band structure (red dotted line) compared with full orbital LDA band structure (black solid line) of FeTe$_2$O$_5$Br}
\end{figure}

\begin{figure}[h]
\centering
\includegraphics[scale=.37]{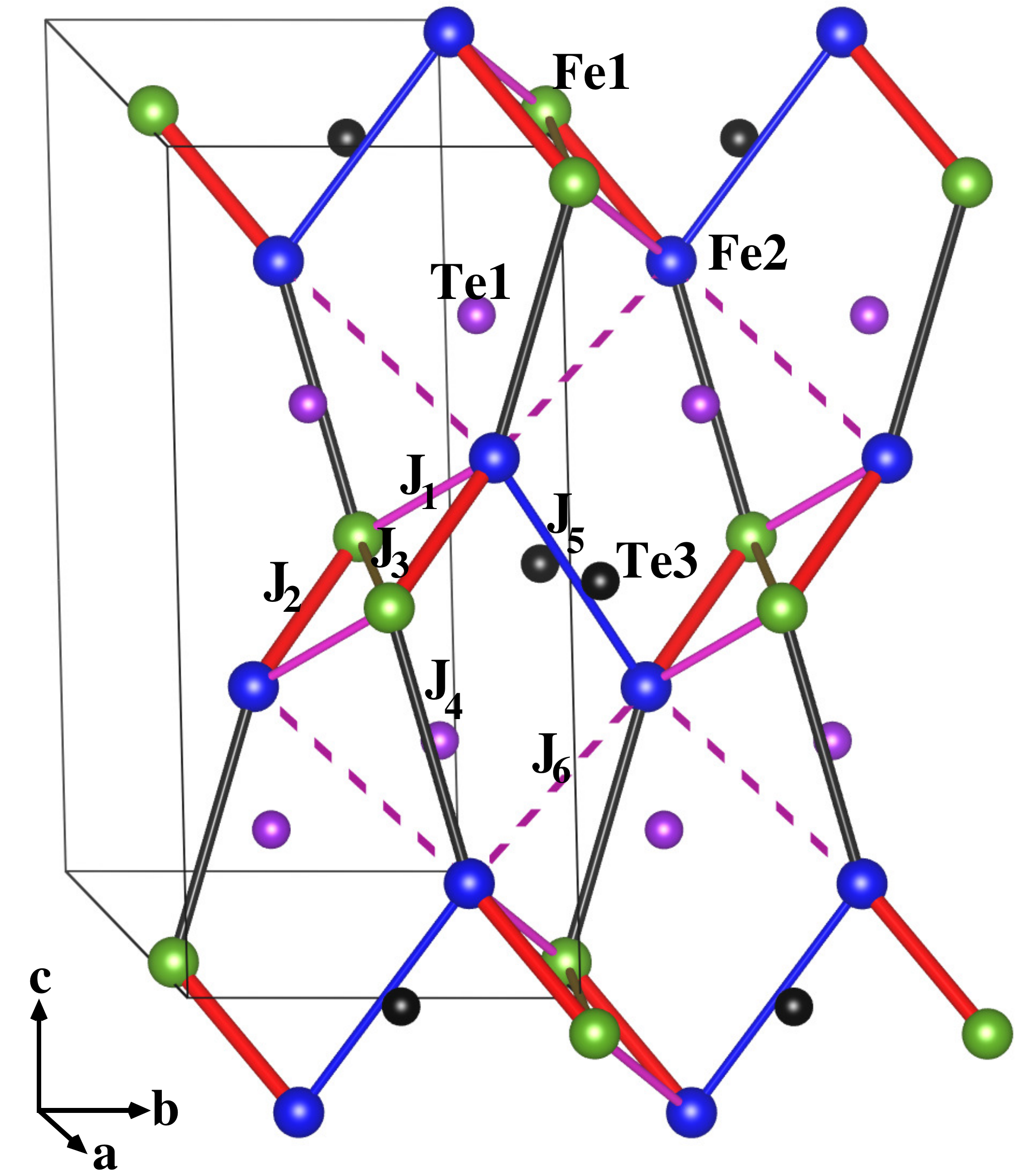}
\caption{\label{fig:jvalue}Structure of FeTe$_2$O$_5$Br, exchange paths are indicated.} 
\end{figure}

Next, we have identified the dominant exchange paths and the relevant spin Hamiltonian using the N$^{th}$  order muffin-tin orbital (NMTO) downfolding method.\cite{Saha-Dasgupta,Tank} In order to derive a low energy effective model Hamiltonian, we have retained the isolated Fe band complex near the Fermi level for a non-spin-polarized calculation and downfolded the rest with the choice of two energy points $E_0$ and $E_1$. The downfolded bands in comparison to the all orbital LDA band structure is shown in Fig.~\ref{fig:nmtoplot} and we note that the agreement is very good. The Fourier transform of the low energy Hamiltonian $\rm {H_k} \rightarrow \rm {H_R}$ [where $\rm {H_R}$ is given by, $H_R=\sum_{ij} t_{ij}\left(c_i^{\dagger}c_j+h.c.\right)$]  gives the effective hopping parameters between the various Fe atoms. The various hopping integrals can be utilized to identify the dominant exchange paths. For strongly correlated systems, the antiferromagnetic contribution to the exchange integral can be computed using  $J^{AFM}$ = 4$\frac{\sum_{i,j}t_{ij}^2}{U_{eff}}$, where $U_{eff}$ is the effective onsite Coulomb interaction  and $t_{ij}$ corresponds to the hopping via superexchange paths. The ratio of the various exchange interactions are displayed in Table~\ref{tab:Jvalue} and the various exchange paths are indicated in Fig.~\ref{fig:jvalue}. In last two columns we have also reproduced  the ratio of exchange interactions obtained in Ref.\onlinecite{henaarxiv} using the total energy method.
%%%----------Fig6......
In Ref. \onlinecite{henaarxiv}, it is reported that the alternating spin chain model is more appropriate instead of the tetramer model  suggested for this system as the inter-tetramer  super exchange ($J_4$) mediated by Fe-O-Te-O-Fe bridges is appreciable. The values of exchange interactions obtained from the NMTO downfolding method reveal that in addition to the intra cluster exchange interactions $J_1$, $J_2$, $J_3$, the inter-cluster exchange interaction  $J_4$ is substantial supporting the validity of alternating spin chain model suggested in Ref. \onlinecite{henaarxiv}. Although the qualitative pictures as given in Ref. \onlinecite{henaarxiv} and in the present study are similar, the quantitative values of the exchange interactions, specifically the values of  $\frac{J_1}{J_2}$ and $\frac{J_4}{J_2}$, differ in the two studies, possibly due to the different calculation schemes adopted in these two independent investigations.
%%%-----------Fig7------------------------------
%\longtable
%\squeezetable
\begin{table*}
\begin{center}
\caption{\label{tab:Jvalue} Exchange interactions along different exchange paths obtained from NMTO downfolding method and energy method (Ref.~\onlinecite{henaarxiv}) have been tabulated here}
\begin{ruledtabular}
\begin{tabular}{ccccccc}
%Types of interaction & Exchange & Distance &  J$_i$/J$_2$ & J$_i$/J$_2$ &J$_i$/J$_2$  \\
% & & (\AA)   & from NMTO method  & in Ref.~\onlinecite{henaarxiv} & in Ref.~\onlinecite{henaarxiv} \\
% & &   (meV) & $U_{eff}=3$ eV &for $U_{eff}=3$ eV &  for $U_{eff}=4$ eV\
 Exchange & Distance & Exchange paths, & J$_i$/J$_2$ & J$_i$/J$_2$ &J$_i$/J$_2$  \\
 & (\AA) &    bond lengths    & from NMTO   & in Ref.~\onlinecite{henaarxiv} & in Ref.~\onlinecite{henaarxiv} \\
 &  & and angles  & ($U_{eff}=3$ eV) &  ($U_{eff}=3$ eV) &  ($U_{eff}=4$ eV)\\
\hline
 $J_{1}$ & 3.16 & $\angle$Fe1-O1-Fe2= 101.8$^\circ$ & 0.89 & 0.46 & 0.35\\
 & & $\angle$ Fe1-O2-Fe2= 99.5$^\circ$  & & & \\
 \hline
 $J_{2}$ & 3.34 &  $\angle$Fe1-O7-Fe2=110.2$^\circ$  &   1 & 1 & 1\\
 & & $\angle$Fe1-O2-Fe2= 95.79$^\circ$ & & & & \\
\hline
 $J_3$ & 3.43 &  $\angle$Fe1-O2-Fe1 = 101.7$^\circ$ &  0.44   & 0.33  & 0.34  \\
\hline
 $J_4$ & 4.76 & Fe1-O-Te1-O-Fe2  & 0.26  & 0.62 & 0.59\\
& & Fe1-O-Te4-O-Fe2  & & &\\
\hline
 $J_5$ & 4.77 & Fe2-O-Te3-O-Fe2 & 0.05  & 0.04 & 0.0 \\
\hline
 $J_6$ & 5.10 & Fe1-O-Te1-O-Fe1 & 0.15   & 0.27 & 0.26\\
\hline
 $J_7$ & 5.52 &  O-O $\sim$ 2.81 & 0.02  & -- & --\\
\end{tabular}
\end{ruledtabular}
\end{center}
\end{table*}

We  next investigated the impact of magnetism on crystal structure, viz., exchange striction. We have carried out  the structure optimization with nonmagnetic, ferromagnetic, and AFM1 magnetic configurations. In this optimization, the cell parameters were fixed to the experimental values, but the positions of the atoms were allowed to relax. The  change in bond lengths  with respect to the unrelaxed (experimental) structure corresponding to  various exchange paths are displayed in Table~\ref{tab:distancerela} for the AFM1, FM and non-spin polarized cases. The bond lengths hardly change due to the ionic relaxations for nonmagnetic and ferromagnetic cases indicating negligible exchange striction.  The maximum change in  ionic positions occur in the relaxed structure with  AFM1 magnetic ordering. The dominant changes  correspond to the exchange path $J_3$ involving oxygens and $J_5$ involving the Te ions (see Table \ref{tab:distancerela}). Our calculations provide a direct evidence that the exchange paths $J_3$ and $J_5$ are responsible for the spin-phonon coupling in this compound. The importance of the exchange path $J_5$ was also anticipated in Ref.~\onlinecite{henaarxiv}.

To obtain an estimate of the impact of structural distortion on the lone pairs, we computed the hybridization index for the relaxed structure in the AFM1 phase. The H-indices for the relaxed structure  are  found to be 17.305 and 11.015 between Te-$p$ and O-$p$, and Te-$s$ and O-$p$ respectively as opposed to 17.00 and 10.99 in the AFM1 phase for the experimental structure.  This indicates that the Te-O hybridization increases as a result of the structural distortion, pointing to the importance of Te lone pairs.  Finally  to access the asymmetry between two neighboring lobe shaped charge distribution of the lone pairs we have calculated the moment of the electron localization function ($\vec{M}_{\text{ELF}}^i$) for the $i^{\text{th}}$ Te atom as follows:
\begin{equation}
        \vec{M}_{\text{ELF}}^i = \int_{|\vec{r}| = 0}^R d^3r \text{ELF}(\vec{r}) \vec{r} \label{eq:MELF},
\end{equation}
where $\vec{r}$ is the position vector assuming the $i^{\text{th}}$ Te atom at the origin and $R$ is a suitably chosen radius of a sphere that covers the range of ELF around the $i^{\text{th}}$ Te atom. We find that the sum of $\vec{M}_{\text{ELF}}^i$'s vanishes for a pair of suitably chosen Te atoms  in the high temperature centrosymmetric experimental structure,\cite{beckerjacs2006}  whereas it has a finite value for the same pairs of atoms in the relaxed structure. This observation suggests that unlike the centrosymmetric experimental structure where the local dipole moments cancel pairwise leading to no net polarization, in the relaxed magnetic structure they do not cancel out. (the average ELF moment for a pair of Te atoms in a relaxed magnetic structure is 7.2 \AA).  This calculation hints at  the activation of the stereochemical activity of the Te ions once the polarization is triggered by the magnetic ordering, as elaborated in the next section. In fact the minute displacements of the Te$^{4+}$ ions below T$_{N_{2}}$ in the multiferroic LT-IC phase has been corroborated by the NQR results. \cite{Pregelj2013} 

%the Te-$5s$ lone pairs however play an important role in inducing polarization in the relaxed structure when the system is magnetic. 

\begingroup
%\longtable
%\squeezetable
\begin{table*}
\begin{center}
\caption{\label{tab:distancerela} The bond lengths  between the Fe atoms in the experimental structure and the change in Fe-Fe bond length upon relaxation within different magnetic configurations have been listed here. $+(-)$ sign indicate increment (decrement) of the distance.}
\begin{ruledtabular}
\begin{tabular}{ccccc}
Exchange  & Bond length (\AA) & \multicolumn{3}{c}{Change in the bond length upon relaxation (\AA)}  \\
paths & Exp. Struc. & \multicolumn{3}{c} {with respect to the experimental structure}\\
 &  &  AFM1 & FM & NM \\
\hline
$J_1$ (Fe1-Fe2) & 3.16 & $-$0.04 & $-$0.01 & $-$0.01\\

$J_2$ (Fe1-Fe2) & 3.34 & $-$0.03 & 0.00 &  0.00 \\

%\cellcolor[gray]{0.8} $J_3$ (Fe1-Fe1)  & \cellcolor[gray]{0.8} 3.43 & \cellcolor[gray]{0.8} $-$0.11 & \cellcolor[gray]{0.8} $-$0.04 & \cellcolor[gray]{0.8} $-$0.02 \\
${\bf {J_3}}$ {\bf{(Fe1-Fe1)}}  & {\bf{3.43}} & {\bf{$-$0.11}} &  {\bf{-0.04}} & {\bf{-0.02}} \\

$J_4$ (Fe1-Fe2)& 4.76 & 0.02 & 0.00 & 0.00\\

%\cellcolor[gray] {0.8} $J_5$ (Fe2-Fe2) & \cellcolor[gray]{0.8}  4.77 & \cellcolor[gray]{0.8}  0.05 & \cellcolor[gray]{0.8}  0.02 & \cellcolor[gray]{0.8}  0.01\\
%\rowcolor{lightgray} {$J_5$ (Fe2-Fe2) & 4.77 & 0.05 & 0.02 & 0.01}\\
${\bf {J_5}}$ {\bf{(Fe2-Fe2)}} & {\bf{4.77}} & {\bf{0.05}} & {\bf{0.02}} & {\bf{0.01}}\\

$J_6$ (Fe2-Fe2) & 5.10 & 0.00 & 0.00 & 0.00 \\

\end{tabular}
\end{ruledtabular}
\end{center}
\end{table*}
\endgroup

\subsection{Polarization}
%=============================================

%{\bf Jayita: Delete the last column in Table IV, for the paper but keep it in your thesis}
We have calculated the ferroelectric polarization  using the Berry phase method\cite{resta1994} as implemented in the Vienna {\em ab-initio} simulation package (VASP).\cite{kresse1993} The polarization calculations are carried out  with the idealized magnetic configuration AFM1 for several $U_{eff}$ values. Our results are summarized in Table \ref{tab:polarization}. The direction of polarization is the same with different $U_{eff}$ values, but the magnitude decreases with the increasing value of  $U_{eff}$. The calculated polarization for FeTe$_2$O$_5$Br is large compared to the experimental value. Such an overestimation is also reported for other systems,\cite{Kan2009,Xiang2007} and may be attributed to the idealized magnetic structure considered in our calculation.  In view of the fact that upon ionic relaxation the bond lengths corresponding to the exchange path $J_3$ and $J_5$ change substantially, we have investigated the impact of the change in bond length on the exchange interaction and hence  on the values of the polarization.   
%%--Fig8---------------
\begin{figure}[h]
\centering 
\includegraphics[scale=.37]{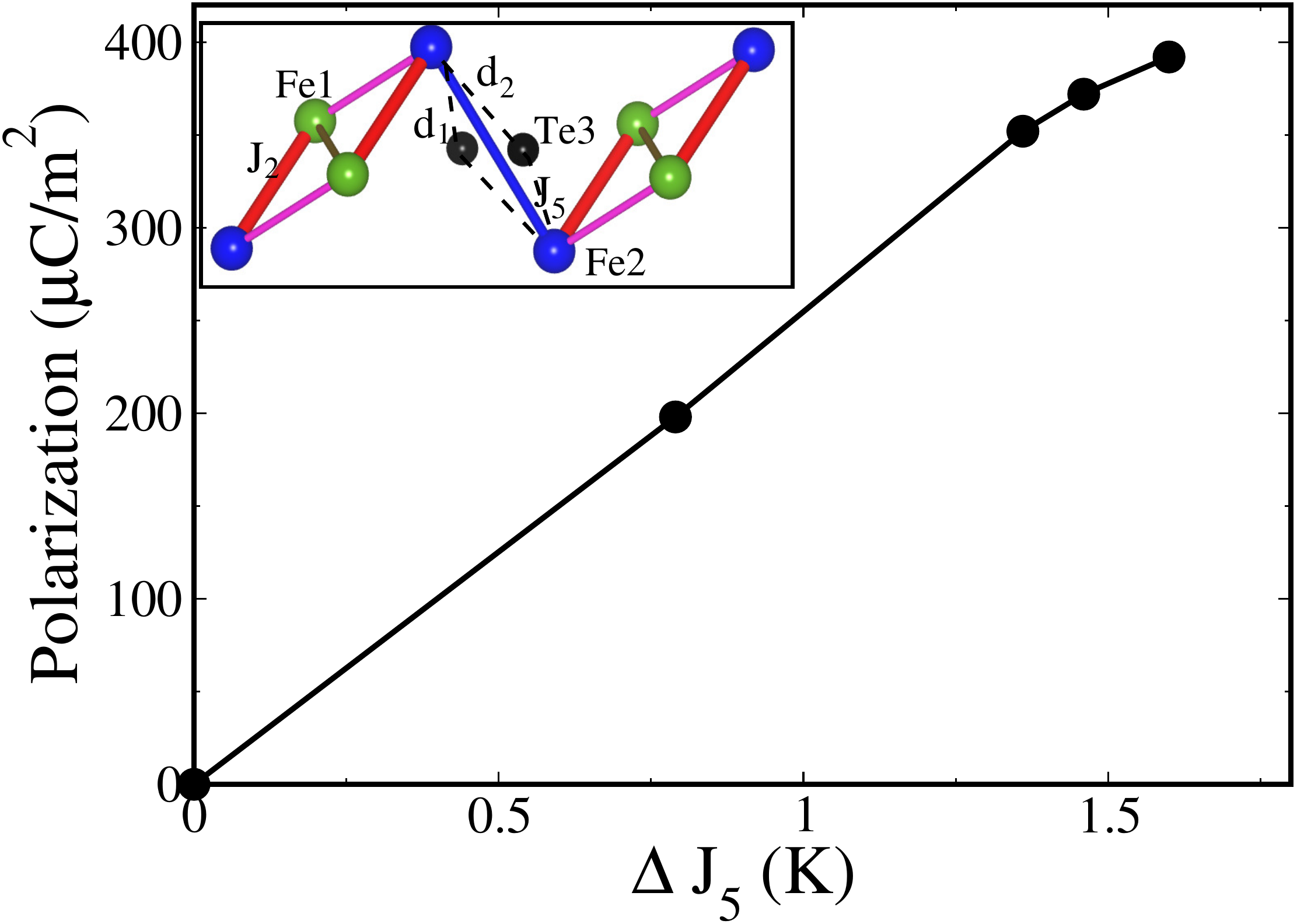}
\caption{\label{fig:tepola}Variation of polarization with $\Delta J_5$. Inset shows the  $J_5$ exchange path involving Te3 ions.}
\end{figure}
The exchange interaction $J_3$  involve  Fe-O-Fe, super exchange pathway and therefore obey the Anderson-Goodenough-Kanamori rules. When the Fe-Fe distance in the $J_3$ exchange path is reduced, not only the $J_3$ increases but also the value of the polarization increases, indicating the importance of this super exchange path on polarization. High resolution synchrotron X-ray diffraction however did not detect significant structural changes for this bond.\cite{henaarxiv}  Next we have investigated the $J_5$ exchange path involving the Te ions. In Ref. \onlinecite{henaarxiv}, it is reported that  the only sizable change in the LT-IC phase corresponds to the shortening of Fe2-Te3 distance in the  $J_5$ exchange pathway.  In order to  see how the displacement of Te3 ions affect the exchange interaction $J_5$ and in turn its effect on the electric polarization, we have changed the distance between Fe2-Te3 ($d_1$) (also the distance $d_2$ between Fe2-Te3) (see inset of Fig.~\ref{fig:tepola}) and computed the exchange interaction $J_5$ and the ferroelectic polarization.   In Fig. \ref{fig:tepola}, we have plotted the polarization as a function of the  change in exchange interaction $\Delta J_5$ (between the distorted and the experimental structure). $\Delta J_5$ may be considered as a  measure of spin phonon interaction mediated by the Te ions. Polarization increases as $\Delta J_5$ is increased and this polarization originates from the spin phonon coupling corresponding to the $J_5$ exchange pathway. Our calculations reveal polarizable lone pairs enhance the spin phonon coupling upon exchange striction in the AFM1 phase which in turn lead to  ferroelectric polarization.  %%%%_____________Polarization with Te positins fixed__________________
In order to check the role of Te ions in the polarization we have carried out constrained ionic relaxation calculation in which positions of Te ions were kept fixed and other ionic positions were allowed to relax for the AFM1 configuration with $U_{eff}=4$ eV.  The value of polarization  is calculated to be 102 $\mu$C/$m^2$, substantially reduced  from the porarization (187.8 $\mu$C/$m^2$) calculated for the relaxed structure where Te ions are also moved from their centrosymmetric positions. This result suggests that exchange striction within the Fe tetramers as well as between them mediated by Te ions, are responsible for the magnetoelectric (ME) effect in FeTe$_2$O$_5$Br.  Interestingly the magnetic ordering also triggers the stereochemical activity of Te ions, giving rise to a feedback mechanism. 
\begin{table}[h]
\begin{center}
\begin{ruledtabular}
\caption{\label{tab:polarization} Calculated electric polarization with AFM1 magnetic configuration with different value of Coulomb interaction parameter $U_{eff}$  for the relaxed structure are listed here.}
\begin{tabular}{cc}
$U_{eff}$ values (eV) & \multicolumn{1}{c}{Polarization ($\mu$C/$m^2$)} \\
\hline
1 & 217.7  \\ 
2 & 208.0  \\
3 & 198.0  \\
4 & 187.8  \\
5 & 177.7  \\
\end{tabular}
\end{ruledtabular}
\end{center}
\end{table}

\section{Conclusions}
We have investigated the electronic properties of a multiferroic compound ${\rm FeTe_2O_5Br}$ by using density functional theory to elucidate the role of Te ions on the ferroelectric polarization of this system.  We find that, in the absence of magnetism the system remains centrosymmetric due to the antipolar orientation of the Te lone pairs that does not promote structural distortion.   The results from our calculations reveal that  FeTe$_{2}$O$_{5}$Br is an improper multiferroic where exchange striction within the Fe tetramers as well as between them is  responsible for the magnetoelectric (ME) effect.
We find that the electric polarization is very sensitive to the $J_5$ exchange path involving the polarizable Te$^{4+}$ lone pairs. %While the lone pairs do not play any role in the absence of magnetism but the lone pairs enhance the spin phonon coupling upon exchange striction which in turn lead to ferroelectric polarization.   
 Te -5$s$ lone pairs show stereochemical activity only when the polarization is triggered by the magnetic ordering.

\section{ACKNOWLEDGMENTS}
I.\ D.\  thanks the Department of Science and Technology, Government of India for financial support, and J.\ C.\ thanks CSIR, India (Grant No. 09/080(0615)/2008-EMR -1) for funding through a research fellowship.

%\bibliography{jayita}

%

\end{document}